\documentclass{article}
\usepackage{amsmath}
\usepackage{amssymb}
\newtheorem{Definition}{Definition}
\newtheorem{Conjecture}{Conjecture}

\newtheorem{Theorem}{Theorem}

 \newcommand{\begeq}{\begin{equation}}
\newcommand{\bea}{\begin{eqnarray}}
\newcommand{\eea}{\end{eqnarray}}

\newcommand{\ca}{$C^*$-algebra} 
 \newcommand{\rep}{representation}

\newcommand{\sthom}{$\mbox{}^*$-homomorphism}
\newcommand{\staut}{$\mbox{}^*$-automorphism}
\newcommand{\stiso}{$\mbox{}^*$-isomorphism}

 \newcommand{\til}{\tilde}
\newcommand{\raw}{\rightarrow} 
\newcommand{\rac}{\rightarrowtail}
\newcommand{\lac}{\leftarrowtail}
\newcommand{\law}{\leftarrow}

\newcommand{\ot}{\otimes} 
 \newcommand{\ra}{\rangle}

\newcommand{\rst}{\upharpoonright} 
\newcommand{\x}{\times} 
 
\newcommand{\cin}{C^{\infty}} \newcommand{\cci}{C^{\infty}_c}

\newcommand{\inv}{^{-1}}

 \newcommand{\Gm}{\Gamma}
 
 \newcommand{\varep}{\varepsilon}

 \newcommand{\phv}{\varphi}

 \newcommand{\CM}{{\mathcal M}}

\newcommand{\CQ}{{\mathcal Q}}

\newcommand{\CZ}{{\mathcal Z}}
\newcommand{\C}{{\mathbb C}} 
 
 \newcommand{\R}{{\mathbb R}}
 
  \makeatletter
\newskip\tempskip \def\endproof{{\parfillskip24\p@ plus\@ne
fil\@@par}\tempskip\prevdepth
\ifdim\lastskip=\z@\tempskip\z@\else\vskip-\lastskip
\ifdim\tempskip>4\p@ \tempskip.5\tempskip \else \tempskip\z@\fi\fi
\nobreak\vskip-\baselineskip\vskip-\tempskip\noindent\hbox
to\hsize{\hfill
$\blacksquare$}\par\vskip\tempskip\vskip\abovedisplayskip\@doendpe}
\makeatother \makeatletter
\newskip\tempskip \def\endiproof{{\parfillskip24\p@ plus\@ne
fil\@@par}\tempskip\prevdepth
\ifdim\lastskip=\z@\tempskip\z@\else\vskip-\lastskip
\ifdim\tempskip>4\p@ \tempskip.5\tempskip \else \tempskip\z@\fi\fi
\nobreak\vskip-\baselineskip\vskip-\tempskip\noindent\hbox
to\hsize{\hfill
$\Box$}\par\vskip\tempskip\vskip\abovedisplayskip\@doendpe}
\makeatother \newcommand{\enp}{\endproof}

\newcommand{\otc}{\circledcirc}
\newcommand{\otg}{\circledast}

\newcommand{\Ca}{\mbox{\textsf{C}$\mbox{}^*$}}

\newcommand{\LG}{\mbox{\textsf{LG}}}

\newcommand{\Po}{\mbox{\textsf{Poisson}}}

\newcommand{\LGt}{\mbox{\textsf{L}$\tilde{\mathsf{G}}$}}

\newcommand{\Ch}{\mathbb{C}[[\hbar]]}
\newcommand{\hh}{[[\hbar]]}
\newcommand{\IC}{\mbox{\rm \textsf{C}$\mbox{}^*$}(I)}
\newcommand{\XC}{\mbox{\rm \textsf{C}$\mbox{}^*$}(X)}
\newcommand{\LPo}{\mbox{\rm \textsf{LPoisson}}}
\newcommand{\KK}{\mbox{\rm \textsf{KK}}}
\newcommand{\RKK}{\mbox{\rm \textsf{RKK}}}
\newcommand{\RKKI}{\mbox{\rm \textsf{RKK}}(I)}
\newcommand{\RKKX}{\mbox{\rm \textsf{RKK}}(X)}
\begin{document}
\pagestyle{plain}
\title{Quantization as a functor\thanks{Based on lectures at the
Workshop on Quantization, Deformations, and New
Homological and Categorical Methods in Mathematical Physics
(Manchester, July 2001)}}
\author{N.P. Landsman\thanks{Korteweg--de Vries Institute for Mathematics,
University of Amsterdam,
Plantage Muidergracht 24,
NL-1018 TV AMSTERDAM, THE NETHERLANDS, 
email \texttt{npl@science.uva.nl}}
\thanks{Supported by a Fellowship from the Royal Netherlands Academy
of Arts and Sciences (KNAW)}}
\date{\today}
\maketitle
\begin{center}
\textbf{Abstract}  
\end{center}
Notwithstanding known obstructions to this idea, 
we formulate an attempt to turn quantization into a functorial procedure. 
We define a category \Po\ of Poisson manifolds, whose objects are
integrable Poisson manifolds and whose arrows are  isomorphism classes 
of regular Weinstein dual pairs;  it follows that identity arrows are 
symplectic groupoids, and that two objects are isomorphic in \Po\ iff
they are Morita equivalent in the sense of P. Xu. It has a subcategory
\LPo\ that has duals of 
integrable Lie algebroids as objects and cotangent bundles as arrows.
We argue that naive \ca ic quantization should be functorial from
\LPo\ to the well-known category \KK, whose objects are separable \ca
s, and whose arrows are Kasparov's KK-groups.  This limited
functoriality of quantization would already imply the Atiyah--Singer
index theorem, as well as its far-reaching generalizations developed
by Connes and others.  In the category \KK, isomorphism of objects
implies isomorphism of K-theory groups, so that the functoriality of
quantization on all of \Po\ would imply that Morita equivalent Poisson
algebras are quantized by \ca s with isomorphic K-theories.  Finally,
we argue that the correct codomain for the possible 
functoriality of quantization is the category \RKK(I), which takes the
deformation aspect of quantization into account.
\newpage
\begin{quote}
``First quantization is a mystery, but second quantization is a functor''
(E. Nelson)
\end{quote}
\begin{quote}
 Comme l'on sait la ``quantification g\'eometrique" consiste \`a 
rechercher un certain foncteur de la cat\'egorie des vari\'etes 
symplectiques et symplectomorphismes dans celle des espaces de 
Hilbert complexes et des transformations unitaires (\ldots)
Il est bien connu qu'un tel foncteur n'existe pas. 
(from A. Crumeyrolle's review \textbf{MR81g:58016} of \cite{Got1})
\end{quote}
 \section{Introduction}
The functoriality of second quantization
(a construction involving exponential Hilbert spaces or Fock spaces)
 mentioned in  our opening quote of Nelson is an almost trivial matter.
  The deep problem suggested by this quote is  the 
possible functoriality of ``first'' quantization, which simply
means the quantization of Poisson manifolds $P$.

The simplest example is probably $P=T^*(\R^n)$ with the
usual Poisson structure. Defining quantization either by the Schr\"{o}dinger
\rep\ $U^S_{\hbar}$ 
of the Heisenberg group $H_n$ in dimension $2n$, or by the Weyl--Moyal prescription
$Q_{\hbar}^W$ (which points of view are essentially equivalent), 
it follows either way that the quantization of $T^*(\R^n)$ is functorial with
respect to affine  linear symplectomorphisms and unitary intertwiners;
 see, e.g.,  \cite{Fol}  or \cite{NPL3}.
Taking Weyl--Moyal quantization to be concrete, this statement 
specifically means that one has
\begin{equation}
Q_{\hbar}^W(f\circ L\inv)=U^M_{\hbar}(L)Q_{\hbar}^W(f)U^M_{\hbar}(L)^*, \label{eq1}
\end{equation}
where $f\in\cci(T^*(\R^n))$ (for simplicity), $L$ is an affine linear
symplectomorphism, and $U^M_{\hbar}$ is the \rep\ of the affine
symplectic group composed of the metaplectic \rep\ of the linear
symplectic group and the (projective) Schr\"{o}dinger \rep\ $U^S_{\hbar}$ of the
translation group in dimension 2n.    As is well known, 
 the Groenewold--Van Hove theorem (cf.\ \cite{Got2,GGT} for
an up-to-date treatment) precludes functoriality under a larger class
of classical transformations \cite{Got1}.  This seems  about all that is known
about the functoriality of (first) quantization. 
  
The  above example has a number of instructive features. Firstly,
$T^*(\R^n)$ has a large amount of symmetry, which is fully exploited
by the Weyl--Moyal quantization prescription.  The rather meager
functoriality properties are a direct consequence of this
symmetry. Indeed, the Berezin--Toeplitz quantization prescription on
$T^*(\R^n)$ (relying on its K\"{a}hler structure), which is physically
as acceptable as the Weyl--Moyal prescription, and is much better
behaved analytically
\cite{NPL3}, enjoys even less functoriality. Since both prescriptions hinge on
rather special properties of $T^*(\R^n)$, for the sake of
generalization it would seem wise not to let the notion of
functoriality of quantization rely on the precise details of a
quantization prescription, but rather on a certain equivalence class to
which it belongs.   

Secondly, the Groenewold--Van Hove no-go theorem suggests that taking unitary
transformations on the quantum side does not leave enough room to manoeuvre
in the codomain category of a potential quantization functor.
 Hence one needs a different class of arrows at least in the quantum category. 
It is convenient to work with \ca s rather than concrete Hilbert spaces;
instead of unitary operators one should then speak of \staut s.
For example, eq.\  (\ref{eq1}) defines
conjugation by $U^M_{\hbar}(L)$ as a \staut\ of the \ca\ of compact operators 
on $L^2(\R^n)$. 

Furthermore, it is extremely unnatural to only work with
simple \ca s (like the compact operators), and  once one has decided to work with 
general \ca s, it goes without saying that one should consider general Poisson
manifolds, instead of merely symplectic ones. The conclusion so far, then, is that
the naive idea that quantization ought to be functorial with respect to isomorphisms 
of Poisson manifolds   and \stiso s of \ca s,
let alone the stronger requirement of functoriality  with respect to
Poisson  maps and \sthom s, respectively, 
has to be given up. 

More suitable categories of \ca s necessarily have
a weaker notion of isomorphism than \stiso.
To obtain powerful results, and also to restore a certain parallel between
the classical and the quantum categories, we will accordingly use a classical
category in which isomorphism of objects is weaker than isomorphism of
Poisson manifolds in the usual sense. 

Our original idea was that one should use Morita equivalence on both sides;
for Poisson manifolds this notion was developed by Xu \cite{Xu2}, whereas 
the older \ca ic theory is due to  Rieffel \cite{Rie1,Rie2}. 
If one has categories in which isomorphism of objects comes down to Morita
equivalence, then the possible functoriality of quantization would imply that
quantization preserves Morita equivalence. Such categories are easily
defined \cite{OBWF,FM}. On the classical side one has the category \Po, whose objects are
integrable Poisson manifolds, and whose arrows are  isomorphism classes 
of regular Weinstein dual pairs.
On the quantum side one has a category
\Ca\  whose  objects are \ca s and whose arrows are unitary equivalence  classes
of Hilbert bimodules (for the latter see also \cite{Sch,Ech}). 

Apart from the (flawed) idea in the previous paragraph, the use of the
categories \Po\ and \Ca\ was in addition motivated by the fact that
the possible functoriality of quantization as a map from \Po\ to \Ca\
would imply the ``quantization commutes with reduction'' principle
(see \cite{NPL3} and references therein). Hence this perhaps somewhat
mysterious principle would appear in a canonical mathematical light.
For let $Q\law S_1\raw P$ and $P\law S_2\raw R$ be regular Weinstein
dual pairs, quantized by an $A$-$ B$ Hilbert bimodule
$\CQ(S_1)$ and a $ B$-$C$  Hilbert bimodule $\CQ(S_2)$,
respectively. Here $B_0\cong C_0(P)$.  Functoriality of quantization
implies $$
\CQ(S_1\otc_P S_2)=\CQ(S_1)\hat{\otimes}_{ B}\CQ(S_2). 
$$
Now composition of arrows $\otc$ in \Po\ is given by symplectic
reduction, whereas the interior tensor product of Rieffel that defines
arrow composition $\hat{\otimes}$ in \Ca\ is a quantized version of
the classical reduction procedure \cite{NPL1,NPL3,OBWF}.  Hence
the left-hand side  is ``quantization after reduction,'' whereas the
right-hand side stands for ``reduction after quantization.''

As will be recalled below, \Po\ has a subcategory \LPo\ whose objects are 
 duals of integrable Lie algebroids, and whose arrows are cotangent bundles.
The point is that there indeed exists a functor from \LPo\ to \Ca\ resembling
quantization on the object side, so that Morita equivalent Poisson manifolds
in \LPo\ are quantized by Morita equivalent \ca s. Although this is a nontrivial
result,  
it was pointed out to the author by Alan Weinstein that it is rather untypical,
and that it would be  mistake to conjecture an extension of this result to all
(integrable) Poisson manifolds 
 (as we did in a previous draft of this paper, of which Weinstein was
the referee). He actually pointed out a class of counterexamples, as follows. 
Take any two tori  of the same (even) dimension, but carrying different symplectic
 structures.
These will always be Morita equivalent as Poisson manifolds \cite{Xu2}. 
However, one can easily choose the symplectic structures in such a way 
that their respective quantizations (as defined in \cite{Rie3,Rie4}) 
fail to be Morita equivalent as \ca s (for any value of $\hbar$); cf.\ \cite{RS}.

These counterexamples show that Morita equivalence is still not coarse
enough on the quantum side to be preserved by quantization, and
suggests that it might be more appropriate to use K-equivalence of \ca
s (i.e., isomorphism of K-groups).  The natural codomain of a possible
quantization functor is then clearly the category \KK, whose objects
are separable \ca s, and whose arrows are Kasparov's KK-groups
\cite{Bla,Kas0}. Although isomorphism of objects in \KK\ is not the same as
isomorphism of their K-groups, the latter is implied by the former,
and the category \KK\ has the enormous computational advantage (for
example, over \Ca) that the Hom-spaces $KK(A,B)$ are abelian groups.
The results in \cite{FM} and \cite{HS} then strongly suggest that
quantization should be  functorial from  \LPo\ to \KK.

It should be mentioned that even this limited functoriality of
quantization would already imply the Atiyah--Singer index theorem as well
as its generalization to foliations due to Connes
\cite{Con80,CS}. Moreover, it further motivates the generalization of
the latter to a general index theorem for Lie groupoids called for in
\cite{LR}. One can only marvel at the  possible implications that the complete
functoriality of quantization would have. 

The use of \KK\ is still unsatisfactory, in that its objects are single
\ca s; one effectively works at some fixed value of Planck's constant $\hbar$ (like in
geometric quantization). In the spirit of deformation
quantization, it is much better to use continuous fields of
 \ca s as the target of the quantization operation, as first proposed by Rieffel
\cite{Rie3}. This suggests the use of the
category \RKK(I) as the codomain of a possible quantization functor.
For technical reasons this category has upper semicontinuous fields
of separable \ca s over the interval $I=[0,1]$ (seen as the parameter space of 
 $\hbar$) as objects, and the so-called representable
KK-groups $\mathcal{R}KK(I,-,-)$ \cite{Kas1} as arrows.
 
The plan of this paper is as follows. In Section \ref{CC} we review the construction
of the ``classical'' category \Po\ \cite{OBWF,FM}, which is the domain of the
alleged quantization functor in any of the approaches we discuss. Section \ref{QC1}
describes the simplest candidate \Ca\ for the codomain category of this functor
\cite{Ech,OBWF,Sch}. This category also lies at the basis of the construction 
of the more sophisticated codomains used further on. In Section \ref{T1}
we prove that quantization is functorial from the subcategory \LPo\ of \Po\ to
\Ca. Section \ref{KK} recalls Kasparov's category \KK\ \cite{Bla,Kas0}, and refines the
previous result so as to apply to \KK\ rather than \Ca. In Section \ref{DQ}
we turn to deformation quantization. In particular, we indicate how the original
ideas of formal deformation quantization \cite{BFFLS} can be realized in the context
of \ca s, so as to motivate both Rieffel's axioms for \ca ic quantization \cite{Rie3}
and the author's modification of these \cite{NPL3}. 
As outlined in Section \ref{RKK}, these considerations immediately
lead to a refinement \IC\ of the category \Ca, in such a way that the step from 
\Ca\ to \KK\ is analogous to the passage from \IC\ to a category \RKK(I).
The latter, then, is our proposal for the codomain of a potential functorial
 quantization procedure. Thus quantization  should be a functor from
\Po\ to \RKK(I). 

Finally, let us note that, in view of the audience towards which these lectures
were directed, some definitions are given in greater detail than others. 
Most participants were familiar with the likes of derived categories and symplectic groupoids,
whereas elementary knowledge of operator algebras seemed lacking.

\bigskip

\textbf{Acknowledgements}
The program in this paper was first presented at the MSRI Workshop on
Quantization and Noncommutative Geometry (Berkeley, April 2001), and
subsequently at the 4th Operator Algebras Conference (Constanza, June
2001), and at the Workshop on Quantization, Deformations, and New
Homological and Categorical Methods in Mathematical Physics
(Manchester, July 2001, present Proceedings).  It is a pleasure to
thank Marc Rieffel, Florian Boca, Birant Ramazan, and Ted Voronov for
their hospitality at these conferences, and to acknowledge many
participants at the above meetings for their comments and
questions. The author is particularly grateful to Kirill Mackenzie for
moral support, and to  Dimitri Shlyakhtenko for critical comments
undoing the latter. 
Subsequently, an earlier draft of this paper (circulated as eprint 
\texttt{math-ph/0107023 v1}, now obsolete) was criticized by Alan
Weinstein, whose comments were instrumental in reaching the point of view
expressed in the present version. 
\section{The classical category} \label{CC}
We recall the definition of the category
of integrable Poisson manifolds introduced in \cite{OBWF,FM}. This category relies on
the theory of symplectic groupoids (cf.\ \cite{CW,CDW} and refs.\ therein, as well
as the forthcoming 2nd edition of \cite{Mac}).
The objects in \Po\ satisfy the following condition.
\begin{Definition}\label{defintP}
A Poisson manifold $P$ is called integrable when there exists a
symplectic groupoid over $P$.
\end{Definition}
This definition is due to \cite{CDW}.  The integrability assumption is
necessary in order to have identities in \Po; see below. In this paper we
assume for simplicity that the symplectic groupoid in question is
Hausdorff. This assumption can be dropped at the expense of
considerable technical complications, which can be overcome if the definition
of  continuous or smooth functions on non-Hausdorff manifolds
in \cite{CM} is used.

The arrows in \Po\ are isomorphism classes of certain Weinstein dual
pairs.  Recall that, given two Poisson manifolds $P$ and $Q$, a
Weinstein dual pair $Q\law S \raw P$, simply called a dual pair in
what follows, consists of a symplectic manifold $S$ and Poisson maps
$q:S\raw Q$ and $p:S\raw P^-$, such that $\{q^* f, p^* g\}=0$ for all
$f\in\cin(Q)$ and $g\in\cin(P)$ \cite{K1,W1}.  Two $Q$-$P$ dual pairs
$Q\stackrel{q_i}{\law}\til{S}_i\stackrel{p_i}{\raw}P$, $i=1,2$, are
isomorphic when there is a symplectomorphism $\phv:\til{S}_1\raw
\til{S}_2$ for which $q_2\phv=q_1$ and $p_2\phv=p_1$.

The notion of regularity for dual pairs is explained in
\cite{OBWF,FM}; its goal is to guarantee the existence of the
following symplectic quotients. Part of the regularity condition is
the stipulation that the maps $p$ and $q$ be complete,
and that $q$ is a surjective submersion.  Let $R$ be a
third integrable Poisson manifold, and let $Q\law S_1\raw P$ and
$P\law S_2\raw R$ be regular dual pairs. The embedding $S_1\x_P
S_2\subset S_1\x S_2$ is coisotropic \cite{NPL3}; we denote the
corresponding symplectic quotient by $S_1\otc_{P}S_2$. This is the
middle space of a regular dual pair $P\law S_1\otc_{P} S_2\raw R$. The
operation $\otc$ is associative up to isomorphism.

For suitable choices of dual pairs, the
product $\otc$ is the same as Marsden--Weinstein reduction
\cite{NPL3}; this should not be surprising in view of its general
definition in terms of symplectic reduction.

Using results in \cite{MM} and \cite{MX}, it can be shown that if $P$
is integrable, then there exists an $s$-connected and $s$-simply
connected symplectic groupoid $\Gm(P)$ whose base space is isomorphic
to $P$ as a Poisson manifold. Moreover, $\Gm(P)$ is unique up to
isomorphism of symplectic groupoids. Cf.\ Lemma 5.6 in \cite{OBWF}.
 The upshot of this is that the isomorphism class 
$[P\stackrel{t}{\law}\Gm(P)\stackrel{s}{\raw}P]$ is a two-sided
identity for $\otc_P$.  (We denote the source
and target maps in a groupoid by $s$ and $t$, respectively.)
\begin{Definition}\label{defPo}
 The category \Po\ has integrable Poisson manifolds as objects, and
 isomorphism classes of regular dual pairs as arrows. 
\end{Definition}

The original reason for the introduction of this category was the fact
that two Poisson manifolds are Morita equivalent in the sense of Xu
\cite{Xu2} iff they are isomorphic objects in \Po; see Prop.\ 5.13 in
\cite{OBWF}.  In particular, a Poisson manifold is integrable iff it
is Morita equivalent to itself (as already observed by Xu).  The category
\Po\ is a classical analogue of the category of \ca s with unitary
equivalence classes of Hilbert bimodules as arrows \cite{OBWF,FM}.

We now introduce a subcategory \LPo\ of \Po\ on which we will be
able to define a quantization functor. This
subcategory is not full, though in an informal sense it is large and
interesting. Recall that a Lie groupoid $G$ over $G_0$ has an
associated Lie algebroid $A(G)$, which is a vector bundle over
$G_0$ \cite{Mac}. The dual vector  bundle $A^*(G)$ is equipped with a canonical Poisson
structure \cite{CDW,Cou} (also cf.\
\cite{CDW,NPL3} for a review).  This Poisson structure is linear, in that the
Poisson bracket of two (fiberwise) linear functions is again linear. Conversely, 
any  linear  Poisson structure is the dual of some Lie algebroid \cite{Cou}
(but this Lie algebroid need not be integrable). Poisson manifolds 
of the form  $A^*(G)$ include all cotangent bundles,
all duals of Lie algebras, all manifolds with zero Poisson bracket, all semidirect
product Poisson structures, and all Poisson manifolds defined by a foliation.

The objects of \LPo\ are the Poisson manifolds $A^*(G)$ associated to
arbitrary Lie groupoids $G$.  The arrows in \LPo\ are isomorphism
classes of regular dual pairs that are of the following form.  Let $G$
and $H$ be Lie groupoids, and suppose that a manifold $M$ is a $G$-$H$
bibundle; we write $G\rac M\lac H$.
This means that $G$ and $H$ act smoothly on $M$ on the left and on
the right, respectively, in such a way that the actions commute; cf.\
\cite{OBWF,Moe,Mrc1,Mrc2}. A construction in
\cite{NPLMRW}, generalizing the momentum map of symplectic geometry
from Lie groups to Lie groupoids, associates a dual pair
\begin{equation} 
A^*(G)\stackrel{J_L}{\longleftarrow} T^*(M)
\stackrel{J_R}{\longrightarrow}A^*(H) \label{gmm}
\end{equation}
 to such a bibundle.  For a
dual pair of this form to be regular, it suffices that the bibundle be principal
\cite{Moe,Mrc1,Mrc2} (also see \cite{OBWF} for a review); 
 this means that the base map $\pi:M\raw G_0$ of the $G$-action on $M$
 is a surjective submersion, and that $H$ acts freely and transitively
 on the fibers of $\pi$.  It follows that $M/H\cong G_0$.  In
 foliation theory principal bibundles are seen as generalized maps
 between leaf spaces (see, e.g., \cite{HS,Mrc1}), and, more generally,
  principal bibundles are sometimes called generalized maps
 between groupoids.

For example, the canonical $G$-$G$ bibundle  $G\rac G\lac G$
gives rise to the dual pair
\begin{equation}
A^*(G)\stackrel{t}{\longleftarrow} T^*(G)
\stackrel{s}{\longrightarrow}A^*(G),\label{iddp}
\end{equation}
where $T^*(G)$ is the cotangent bundle  defined in \cite{CDW}.
This is precisely the symplectic groupoid associated to
the Poisson manifold $A^*(G)$.

Let \LG\ be the category of Lie groupoids, whose arrows are
isomorphism classes of principal bibundles (see
\cite{OBWF,Mrc1,Mrc2}). Composition of arrows is defined as follows.
Suppose one has  principal bibundles $G\rac M\lac H$ and $H\rac
N\lac K$.  The fiber product $M\times_{H_0} N$ carries a right $H$
action, given by $h:(m,n)\mapsto (mh,h\inv n)$ (defined as
appropriate). The orbit space 
\begin{equation}
 M\otg_H N=(M\times_H N)/H \label{123}
\end{equation}
 is a
$G$-$K$ bibundle in the obvious way. This defines a product on matched
bibundles, which becomes associative on isomorphism classes.  We
define \LGt\ as the full subcategory of \LG\ whose objects are
$s$-connected and $s$-simply connected Lie groupoids.

According to Thm.\ 3 and eq.\ (4.30) in \cite{FM}, 
the above procedure defines a functor $A^*$ from  \LGt\  to \Po. That is,  on
objects one has $G\mapsto A^*(G)$, whereas on arrows
the functor in question maps the isomorphism class of a $G$-$H$ bibundle  $M$ 
to the isomorphism class of the dual pair (\ref{gmm}). The operation $A^*$
may also be defined on \LG,  but it fails to be functorial  because identities are
not always preserved.  
Note that $A^*$ indeed maps \LGt\ into \Po: $A^*(G)$ is actually  integrable,
with  associated symplectic groupoid $T^*(G)$.
\begin{Definition}\label{defPoL}
The category \LPo\ is the image of the functor
\end{Definition} 
$$A^*:\LGt\raw\Po.$$
 
Thus \LPo\  has Poisson manifolds of the form $A^*(G)$, where $G$ is a Lie
groupoid, as objects, and isomorphism classes of cotangent bundles  of the
form (\ref{gmm}), where $M$ is a principal bibundle, as arrows. Note that
\LPo\ contains all identities as appropriate, since 
the  symplectic groupoid $T^*(G)$ is $s$-connected and $s$-simply connected
whenever $G$ is.   
\section{The simplest quantum category}\label{QC1}
Within the Hilbert space formalism for quantum mechanics, it is natural to
assume that the observables of a quantum system form a \ca\ \cite{Haa,NPL3}. 
 Recall that a \ca\ $A$ is a complex associative algebra with involution,
equipped with the structure of a Banach space, such that $\| ab\|\leq \| a\| \|b\|$
and $\| a^*a\|=\| a\|^2$ for all $a,b\in A$. A \ca\ can always be 
faithfully represented as a norm-closed involutive
 algebra of bounded operators on a Hilbert space, on which the involution 
is just the adjoint, and the norm is the usual operator norm. See \cite{Con}
for an overview of the use of \ca s in modern mathematics. 

At some fixed value of Planck's constant $\hbar$, quantization then associates
a \ca\ to a Poisson algebra. This view of quantization is not satisfactory,
and ought to be replaced by the idea of deformation quantization, but for the 
moment we stick to it for pedagogical reasons. Thus, to a first approximation,
 the objects of the codomain category of a possible quantization functor
should be \ca s. In order to assemble these into a category, the most obvious choice
would be to take the arrows to be \sthom s, but the pertinent isomorphisms would
then be \stiso s. 
As mentioned in the Introduction, this choice is inappropriate for quantization theory.
A more suitable class of arrows between \ca s is formed by (isomorphism classes
of) so-called Hilbert bimodules. 

A Hilbert bimodule is the \ca ic analogue of a bimodule for algebras
over a given ring \cite{Rie1,Rie2}. A new feature compared to the
purely algebraic situation is that an $A$-$B$ Hilbert bimodule is
endowed with a $B$-valued inner product.  The complete definition is
as follows.
\begin{Definition}
Let $A$ and $B$ be \ca s.  An $A$-$B$ Hilbert bimodule is an $A$-$B$
bimodule $E$ (where $A$ and $B$ are seen as complex algebras, so that
$E$ is a complex linear space) with a compatible $B$-valued inner
product. Thus there is a sesquilinear map $\langle \,,\,\ra :E\x E\raw
B$, linear in the second and antilinear in the first entry, satisfying
$\langle x,y \ra^* =
\langle y, x\ra$, $\langle x, x \ra \geq 0$,
and $\langle x,x\ra = 0$ iff $x=0$.  The compatibility of the
inner product with the remaining structures means that firstly $E$ has
to be complete in the norm $\| x\|^2= \|\langle x,x\ra \|$, secondly
that $\langle x,yb\ra =\langle x,y\ra  b$, and thirdly that
$\langle a^* x,y\ra=\langle x,ay\ra$ for all $x,y\in E$, $b\in B$, and $a\in A$.
Finally, the left action of $A$ on $E$ is required to be nondegenerate in the
sense that $AE$ is dense in $E$. 
\end{Definition}

Note that $A$-$\mathbb{C}$ Hilbert bimodules are just Hilbert spaces
carrying a nondegenerate \rep\ of $A$. On the other hand, a $\C$-$B$
Hilbert bimodule under the obvious action of $\C$ (by multiples of the
unit operator) is called a Hilbert $B$ module.

 The basic example of an $A$-$A$ Hilbert
bimodule is $E=A$ with the obvious actions and the inner product
$\langle a,b\ra=a^*b$. See \cite{Lance} for the basic theory, cf.\
\cite{OBWF} for a comparison between Hilbert bimodules and analogous
structures in mathematics, and have a look at \cite{NPL3} for the use
of Hilbert bimodule in quantization theory.  One feature of algebraic
bimodules that survives in the Hilbert case is the existence of a
bimodule tensor product \cite{Rie1,Rie2}: from an $A$-$B$ Hilbert
bimodule $E$ and a $B$-$C$ Hilbert bimodule $\til{E}$ one can form an
$A$-$C$ Hilbert bimodule $E\hat{\ot}_B \til{E}$, called the interior tensor
product of $E$ and $\til{E}$. For the following definition \cite{Ech,OBWF,Sch}
we also need the notion of unitary equivalence, which the reader may
guess (see \cite{Lance}, p.\ 24).
\begin{Definition}\label{defCa}
The category \Ca\ has $C^*$-algebras as objects, and unitary equivalence 
class\-es of Hilbert bimodules as arrows. The arrows are composed by
the  interior tensor product.  
\end{Definition}
 
 It follows that the identity arrow at $A$ is the canonical Hilbert bimodule $A$
defined above. In addition, it turns out that two \ca s are Morita equivalent
\cite{Rie2} iff they are isomorphic as objects in \Ca\  \cite{Ech,OBWF,Sch}.
This property suggests that the category \Ca\ should be regarded as a
quantum version of \Po. 
\section{Functoriality of simple quantization} \label{T1}
We are now in a position to state the first   result on the functoriality of quantization. 
\begin{Theorem}\label{main}
There exists a functor $\CQ: \LPo\raw \Ca$ that on objects maps $A^*(G)$ to $C^*(G)$.
\end{Theorem}

The reason why $C^*(G)$ may indeed be seen as the quantization of
$A^*(G)$ at some fixed value of $\hbar$ is actually to be found in
deformation quantization \cite{NPL3,LGCA,LR,Ram}.

\textit{Proof.} 
The functor $\CQ$ is the composition of the following functors:
\begin{equation}
\LPo\stackrel{(A^*)\inv}{\longrightarrow}\LGt  
\stackrel{C^*}{\longrightarrow}\IC. \label{dia}
\end{equation}
We discuss the functors $(A^*)\inv$  and $C^*$ in turn.

Firstly, it follows from Props.\ 3.3 and 3.5 in \cite{MM} that
for $s$-connected and $s$-simply connected Lie groupoids $G$ the
association $G\mapsto A^*(G)$ is invertible. Hence
$A^*:\LGt\raw\LPo$ is an isomorphism of categories, with inverse $(A^*)\inv$.
(It would have been sufficient for $A^*:\LGt\raw\LPo$ to define an equivalence
of  categories, for even then it would possess an inverse up to natural isomorphism, which
would be enough for our purposes.) 

Secondly, on objects the map $G\mapsto C^*(G)$ from Lie groupoids to
\ca s is the well-known association of a convolution \ca\ to a Lie
groupoid \cite{Con} (or, more generally, to a locally compact groupoid
with Haar system \cite{Ren}).  Following a special case in \cite{MRW}
(in which the arrows were taken to be Morita equivalences of
groupoids), the map $G\mapsto C^*(G)$ was extended to a functor from
\LG\ to the category \Ca\   in \cite{FM}.   

It follows that $\CQ=C^*\circ (A^*)\inv:\LPo\raw\Ca$ is a functor.  
\enp
 
We can illustrate this result by noting that an identity (\ref{iddp})
in \LPo\ is mapped to the canonical $C^*(G)$-$C^*(G)$ Hilbert bimodule
$C^*(G)$. Hence the symplectic manifold $T^*(G)$ and the Poisson manifold
$A^*(G)$ are both mapped into the \ca\ $C^*(G)$, but $T^*(G)$ is seen
as an arrow, and $A^*(G)$ is regarded as an object in \LPo, so that
the former is mapped to $C^*(G)$ as (the middle space of) 
a Hilbert bimodule, whereas the latter is sent to $C^*(G)$ seen as
a \ca.
\section{The category \KK}\label{KK}
As mentioned in the Introduction, the preceding theorem cannot be
extended to all of \Po, because noncommutative tori provide
counterexamples.  We therefore propose to replace the category \Ca\ by
an analogous category \KK, in which at least these counterexamples are
circumvented, and whose use is very attractive in many ways. Perhaps
the main motivation for looking at quantization as a functor taking
values in \KK\ is that the well-known relationship between quantum
mechanics and index theory \cite{ENN,Fed,Vor} would be clarified
by such functoriality.

The category \KK\ emerged from Kasparov's work on K-theory \cite{Kas0},
and is discussed in detail in \cite{Bla}. Here we only sketch the main points
that allow one to understand that \KK\ is a subtle and deep modification of \Ca.
Given two separable \ca s $A$ and $B$, one defines $\mathbb{E}(A,B)$
as the collection of $A$-$B$ Hilbert bimodules $E$
that are  countably generated in $B$, and are equipped with the following
 additional structure.  

Firstly,
$E$ should be of the form $E=E_1\oplus E_2$, where each $E_i$ is an
$A$-$B$ Hilbert bimodule. Secondly, and this is the main feature,
there should be an  operator $F:E\raw E$ that is adjointable  (i.e., there is
$F^*:E\raw E$ such that $\langle F^* x,y\ra=\langle x,Fy\ra$) and 
odd (in that $F(E_1)\subseteq E_2$ and $F(E_2)\subseteq E_1$).
This $F$ should be an almost unitary intertwiner of $A\rst E_1$ and
$A\rst E_2$, in that for each $a\in A$ the operators $[F,a]$, 
$(F^2-1) a$, and $(F-F^*)a$ be compact. (Here an 
operator on an $A$-$B$ Hilbert bimodule $E$ is said to be compact
when it can be approximated in norm by linear combinations of
rank one operators of the form $z\mapsto x \langle y,z\ra$ for $x,y\in E$. 
In noncommutative geometry, compact operators are treated as infinitesimals
\cite{Con}.)

This setting was mainly motivated by index theory, in which $A=C(X)$
for some compact manifold $X$, $B=\C$, $E$ is the Hilbert space of
$L^2$-sections of some spinor bundle over $X$, and $F$ is a
pseudodifferential operator; see\cite{HR}. In general, the category
\Ca\ is enormously enriched by requiring the presence of $F$. However,
when the $A$ action on some $A$-$B$ Hilbert bimodule $E_1$ happens to
be by  compact operators, one may choose $E=E_1\oplus 0$ and $F=0$
so as to obtain an element of $\mathbb{E}(A,B)$. Such elements may
even survive the step to be explained next.

Elements $(E,F)$ of  $\mathbb{E}(A,B)$ are called Kasparov cycles. Elements of
$KK(A,B)$ are equivalence classes of such cycles under the following
relations: unitary equivalence,  translation of $F$ along norm-continuous path,
and the addition of degenerate Kasparov cycles. The latter are those
for which the operators $[F,a]$, $(F^2-1) a$, and $(F-F^*)a$ are 0 for all 
$a$. The ensuing equivalence relation may be reexpressed in a number
of alternative forms \cite{Bla}. 

It is not difficult to see that $KK(A,B)$ is an abelian group; the group operation
is the direct sum of both bimodules and operators $F$, and the inverse of
the class of a Kasparov cycle $(E,F)$ is the class of $(E^{\mathrm{op}},-F)$
(where $E^{\mathrm{op}}$ is $E$ with the opposite grading).
Morover, the association $(A,B) \mapsto KK(A,B)$ is contravariantly functorial
in the first entry, and covariantly functorial in the second. One recovers
K-theory as a special case of KK-theory by $K_0(A)\cong KK(\C,A)$,
with topological K-theory as the special case $K^0(X)\cong K_0(C(X))$
whereas K-homology \cite{HR} emerges as $K^0(A)\cong KK(A,\C)$.

The deepest aspect of Kasparov's theory is the existence of the so-called
intersection product
$$
KK(A,B)\x KK(B,C)\raw KK(A,C),$$
which is functorial in all conceivable ways. This leads to the category \KK,
whose objects are separable \ca s, and whose arrows are the KK-groups.
More precisely, the Hom-space of arrows from $B$ to $A$ is $KK(A,B)$.
In addition, $KK(A,B)$ defines a space of homomorphisms from
$K_0(A)$ to $K_0(B)$ through the intersection product
$$
KK(\C,A)\x KK(A,B)\raw KK(\C,B)\cong K_0(B).$$
In particular, if two \ca s are isomorphic in \KK, then their K-groups
are isomorphic.

Refining Theorem \ref{main}, we now conjecture that there exists a
functor from \Po\ to \KK\ that on objects maps $A^*(G)$ to $C^*(G)$.
The evidence for this idea comes from the noncommutative geometry
approach to index theory \cite{Con80,Con,CS}, as follows. The crucial
step in the proof of the K-theoretic version of the Atiyah--Singer
index theorem \cite{AS1} is the association of a Gysin or wrong-way
map $f!: K^0(X)\raw K^0(Y)$ to a continuous (and usually smooth) map
$f:X\raw Y$ between locally compact spaces (usually manifolds) $X$ and
$Y$. For example, an embedding $M\hookrightarrow \R^n$ with pullback
$T^*(M)\hookrightarrow \R^{2n}$ induces a map $K^0(T^*(M))\raw 
K^0(\R^{2n})\cong\mathbb{Z}$, which is the topological index of Atiyah
and Singer \cite{AS1}. In KK-theory, $f!$ is seen as an element of
$KK(C_0(X),C_0(Y))$, inducing the map between $K^0(X)$ and $K^0(Y)$
through the intersection product  as explained above. The proof of the index
theorem hinges on the property
\begin{equation}
(g\circ f)! =f! g!, \label{ConSka}
\end{equation}
where the right-hand side is given by the intersection product.

To generalize this, it is useful to regard a space $X$ as a groupoid
(in which $X$ is both the base and the total space of the groupoid,
and the source, target, and object inclusion maps are all equal to the
identity map), so that $C_0(X)$ is the \ca\ $C^*(X)$ of the groupoid
$X$. One may then attempt to generalize the setting of the preceding
paragraph to construct a functor from \LG\ to \KK. In other words, an
object $G$ is mapped into $C^*(G)$, and a principal $G$-$H$ bibundle,
which we now call $F$ with some abuse of notation, is mapped into
$f!\in KK(C^*(G),C^*(H))$. Then (\ref{ConSka}) is to hold, along with
the preservation of identities.

For the longitudinal index theorem for foliations \cite{Con80,Con,CS}
it is sufficient to do this for the case that $G$ is a space $X$ and $H$ is the
holonomy groupoid of a foliation. The more symmetric case that
$G$ and $H$ are both holonomy groupoids was treated  in \cite{HS};
in both situations one has to impose an additional technical condition
(of K-orientability) on $F$. The case that $G$ and $H$ are both arbitrary
Lie groupoids has not been treated yet in the literature, but this
should be possible. The ensuing functor from \LG\ to \KK\ could then
be composed with the functor $(A^*)\inv$ from the proof of Theorem \ref{main}
to obtain the desired functor from \LPo\ to \KK.  Our conjecture then asks for
an extension of this functor from \LPo\ to \Po.
\section{From formal to \ca ic deformation quantization} \label{DQ}
As mentioned in the Introduction,
the use of the category \KK\ is still unsatisfactory.  Its objects are single
\ca s, describing quantum-mechanical algebras of observables 
at some fixed value of $\hbar$. However, in the context of quantization theory
it is important to study quantum theory for a range of values of Planck's
``constant'' $\hbar$, and to control the classical limit. 
This can be done in a purely algebraic way \cite{BFFLS}, or in an
analytic \ca ic way, as first proposed by Rieffel  \cite{Rie3}  (also cf.\ \cite{NPL3}).

We start with some remarks on the purely algebraic approach, called
formal deformation quantization or star-product quantization,
 which serve the purpose of stressing the analogy between
formal and \ca ic deformation quantization.

A star-product on a Poisson manifold $P$ endows the free module
$$\cin(P)\hh=\cin(P,\C)\otimes_{\C}\Ch $$
over the commutative ring $\Ch$ of complex formal power series
in one variable with the structure of an associative unital algebra over $\Ch$
 (whose product is conventionally written as $*$), in such a way that
$$\cin(P)\hh/\hbar\cin(P)\hh\cong \cin(P)$$ as algebras over $\C$ (so that
$f*g-g*f=0$ in $\cin(P)\hh/\hbar\cin(P)\hh$), and that Dirac's condition
$$ f*g-g*f +i\hbar \{f,g\}=0$$  holds in $\cin(P)\hh/\hbar^2\cin(P)\hh$. 
To state these axioms, it is crucial that there is a canonical map
$f\mapsto f=f+0\cdot\hbar + 0\cdot\hbar^2+\cdots$ from $\cin(P)$ to
$\cin(P)\hh$. Now a unital algebra $A$ over $\Ch$ is nothing but a $\Ch$-algebra
$A$ in the sense that there is an injective  ring homomorphism from $\Ch$ into the center
of $A$; cf.\ \cite[p.\ 121]{Lang}. 
Hence, one could define a generalized star-product 
 on a Poisson manifold $P$ as an associative unital  $\Ch$-algebra
$A$ such that 
\begin{enumerate}
\item $A/\hbar A\cong \cin(P)$ as algebras over $\C$;
\item  there is a cross-section
$Q:\cin(P)\raw A$ of the canonical projection $\pi:A\raw A/\hbar A$ 
for which Dirac's condition holds in the sense that
$$Q(f)*Q(g)-Q(g)*Q(f)+i\hbar Q\left(\{f,g\}\right)=0$$ in $A/\hbar^2 A$. 
\end{enumerate}

Rieffel's  analytic approach \cite{Rie3}, based on the use of continuous fields
of \ca s, was a direct analogue of the original definition
of a star-product, in that his fiber algebras $A_{\hbar}$ were obtained
by putting an $\hbar$-dependent product $*_{\hbar}$ as well as an $\hbar$-dependent norm
$\|\cdot \|_{\hbar}$ on $\cin(P)$ (assuming, for simplicity, that $P$ is compact), and completing.
Hence also here one has a canonical map 
$f\mapsto f$, this time from $\cin(P)$ to $A_{\hbar}$ (for each value of $\hbar$), 
in terms of which Rieffel formulated Dirac's condition as
$$
\lim_{\hbar\rightarrow 0} \| \frac{i}{\hbar}(f*_{\hbar} g-g*_{\hbar}f) -\{f,g\}\|_{\hbar} =0.
$$
It was subsequently realized that more general continuous fields of \ca s were needed in 
order to incorporate examples related to Berezin--Toeplitz quantization; cf.\  \cite{NPL3} 
and references therein. In the present context, such fields are best described using
the formalism of $C(X)$ \ca s, which we now recall.

The following definition is due to Kasparov \cite{Kas1} (in the more general
case of locally compact $X$). We will only need the case $X=I$.
\begin{Definition}
Let $X$ be a compact Hausdorff space. A $C(X)$ \ca\   is a \ca\ $A$
with a unital embedding of $C(X)$ in the center of its multiplier algebra.
In other words,  $A$ comes equipped with a unital injective \sthom\  
$C(X)\raw \CZ(\CM( A))$.
\end{Definition}

The structure of $C(X)$ \ca s was fully clarified by Nilsen \cite{Nil}, as follows.
A field of \ca s is a triple $(X, \{ A_x\}_{x\in X},  A)$,
where $\{ A_x\}_{x\in X}$ is some family of \ca s indexed by $X$, and
$A$ is a family of sections (that is, maps $f:X\raw \coprod_{x\in X} A_x$ for which
$f(x)\in A_x$) that is 
\begin{enumerate}
\item
 a \ca\ under pointwise operations and the natural norm 
$$\| f\|=\sup_{x\in X} \| f(x)\|_{ A_x};$$
\item 
 closed under multiplication by $C(X)$;
\item
full, in that for each $x\in X$ one has $\{f(x)\mid f\in\Gm\}= A_x$.
\end{enumerate}
The field is said to be continuous when for each $f\in A$ the function $x\mapsto \| f(x)\|$
is in $C(X)$ (this is equivalent to  the corresponding definition of Dixmier \cite{Dix}; cf.\ 
\cite{KW}). 
 The field is upper semicontinuous when for each $f\in A$
and each $\varep>0$ the set $\{x\in X\mid \| f(x)\|\geq\varep\}$ is compact.

Thm.\ 2.3 in \cite{Nil} now states that  a $C(X)$ \ca\  $A$ defines a unique
upper semicontinuous field of \ca s $$(X, \{ A_x= A/C(X,x) A\}_{x\in X},  A).$$
Here $$C(X,x)=\{f\in C(X)\mid f(x)=0\},$$
and, with  abuse of notation, $f\in A$ is identified with the section
 $$x \mapsto \pi_x(f),$$ where $\pi_x: A\raw  A_x$ is the canonical projection.

Moreover, Blanchard \cite{Bla1}
proved that  a $C(X)$ \ca\  $A$ defines a continuous field of \ca s whenever
the map $x\mapsto \| \pi_x(f)\|$ is continuous for each $f\in A$. 
Thus a continuous field of \ca s over $X$ may be described as a $C(X)$ \ca\ 
with this additional continuity condition.

It should be noted that, unlike vector bundles,  
continuous fields of \ca s may well fail to be
locally trivial. Here the restriction of a continuous field
$(X, \{ A_x\}_{x\in X},  A)$ to some open subset $Y\subset X$ is
said to be trivial when $A_x=B$ for all $x\in Y$, and $A$ contains
$C_0(Y,B)$. In \ca ic deformation quantization, where $X=I$,
both the situation that the field is trivial at $(0,1]$
and the case that all fiber algebras $A_{\hbar}$ are pairwise non-isomorphic
occur! The former happens, for example, in Weyl--Moyal quantization and
its generalizations, whereas the latter occurs for certain noncommutative tori.
 
In any case, we see that a continuous field of
\ca s over the interval $I$ is nothing but a $C(I)$ \ca\ $A$ with an additional
continuity property. Hence, in analogy with  the notion of a generalized
star-product introduced above,  we may reformulate
Def.\ II.1.2.5 in \cite{NPL3} as follows. 
\begin{Definition}\label{gsq}
A strict quantization of a Poisson manifold $P$ is a 
$C(I)$ \ca\ $A$ such that
\begin{enumerate}
\item
$A_0= A/C(I,0) A\cong C_0(P)$ as \ca s;
\item There exists a cross-section $Q$ of $\pi_0$, defined on a suitable Poisson subalgebra of
$C_0(P)$, such that, in terms of  $Q_{\hbar}=\pi_{\hbar}\circ Q$, 
$$
\lim_{\hbar\rightarrow 0} \|
\frac{i}{\hbar}\left(Q_{\hbar}(f)Q_{\hbar}(g)-Q_{\hbar}(g)Q_{\hbar}(f)\right)-Q_{\hbar}(\{f,g\})
\| =0;
$$
\item For each $A\in A$, the function $\hbar\mapsto \| \pi_{\hbar}(A)\|$ from $I$ to $\R^+$
is  continuous.   
\end{enumerate}
\end{Definition}
Here the norm and the product  are taken in $A_{\hbar}$.

The naive quantization of $A^*(G)$ by $C^*(G)$ can be amplified into a strict quantization
\cite{NPL3,LGCA,LR,Ram}. The $C(I)$ \ca\ quantizing $A^*(G)$ turns out to be
$C^*(G_T)$, where $G_T$ is the generalized tangent groupoid of $G$ \cite{HS}.
 \section{The category \RKK}\label{RKK}
We now define categories \IC\ and \RKK, which are the appropriate substitutes of
\Ca\ and \KK, respectively, if one works with 
 $C(I)$ \ca s rather than merely with \ca s. Our goal being  a quantization functor
with codomain \RKK, for pedagogical reasons we first  introduce \IC. 

We first generalize a definition of Blanchard \cite{Bla2}, who
considered the case $ B=C(X)$ (also cf.\ \cite{LG}).
\begin{Definition}\label{Legall}
Let $A$ and $ B$ be $C(X)$ \ca s. An $A$-$ B$ $C(X)$ Hilbert bimodule is an
$A$-$ B$ Hilbert bimodule for which the $A$ action is $C(X)$-linear.
\end{Definition}

The $C(X)$-linearity means the following: since the left action of $A$
on $E$ and the right action of $ B$ on $E$ are both nondegenerate,
they extend to the respective multiplier algebras, so that a priori
one obtains two different actions of $C(X)$ on $E$, coming from $A$
and $ B$ seen as $C(X)$ \ca s.  These actions must
coincide. Consequently, one obtains a field $(E_x)_{x\in X}$ of
$A_x$-$ B_x$ Hilbert bimodules, where $$E_x=E\hat{\otimes}_{ B} B_x$$
is the interior tensor product of $E$ (as an $A$-$ B$ Hilbert
bimodule) and $ B_x$ (as a $ B$-$ B_x$ Hilbert bimodule). The left
action of $ B$ on $ B_x$ is defined through $\pi_x: B\raw B_x$ and
left multiplication, the right action of $ B_x$ on itself is given by
right multiplication, and the $ B_x$-valued inner product on $ B_x$ is
$\langle A,B\rangle=A^*B$ as usual.  The left action of $A_x$ on
$E_x$ is well defined because of the $C(X)$-linearity of the given
$A$ action on $E$.  Thus one may think of an $A$-$ B$ $C(X)$ Hilbert
bimodule as a field of $A_x$-$ B_x$ Hilbert bimodules with certain
continuity properties following from the above definition; the special
case $ B=C(X)$ (so that $ B_x=\C$) considered in \cite{Bla2}
corresponds to a field of nondegenerate \rep s of $A_x$ on a field of
Hilbert spaces over $X$.
\begin{Definition}\label{def4}
Let $X$ be a compact Hausdorff space. 
The objects of the category \XC\ are $C(X)$ \ca s. The arrows are
unitary isomorphism classes of $C(X)$ Hilbert bimodules.
Matched arrows are composed through Rieffel's interior tensor product.
 The identity arrow $1_{ A}$ at an object $A$ is the class
of $A$,  seen as the canonical $A$-$A$ Hilbert bimodule.
\end{Definition}

When $X$ is a point, one recovers the category \Ca.  When $X=I$, one
has the category \IC.  The categories \XC\ are the appropriate \ca ic analogues of
the categories $k$- Alg\ in the purely algebraic setting, where $k$ is
a commutative ring (cf.\ \cite{OBWF}, Sect.\ 2.1); the objects of $k$-
Alg\ are associative unital algebras over $k$, and the arrows are
isomorphism classes of bimodules, composed using the obvious tensor
product.  In particular, \IC\ is the \ca ic counterpart of the
category $\Ch$- Alg. It can be shown that
 two $C(X)$ \ca s are isomorphic as objects in \XC\ iff they are
Morita equivalent as $C(X)$ \ca s (this means that they
are Morita equivalent through 
 an imprimitivity bimodule \cite{Rie1} that is also
a $C(X)$  Hilbert bimodule). 
  The purely algebraic counterpart of this result is
that two objects are isomorphic in $k$- Alg\ iff they are Morita equivalent (in the
usual algebraic sense); see Prop.\ 2.4 in \cite{OBWF}, and also cf.\ \cite{BW}.
 
We now define the category \RKKI, which refines \IC\ in the same way that
\KK\ refines \Ca. In fact, one may define a category \RKKX\ for any
compact Hausdorff space $X$ \cite{PT}.  As with \KK, one starts with
the notion of a Kasparov cycle.  Given separable $C(X)$ \ca s $A$ and
$B$, the elements of $\mathcal{R}\mathbb{E}(X;A,B)$ are those elements
$(E,F)\in \mathbb{E}(A,B)$ for which $E$ is an $A$-$B$ $C(X)$ Hilbert
bimodule. There is no additional condition on $F$.  The group
$\mathcal{R}KK(X;A,B)$ is then defined in precisely the same way as
$KK(A,B)$, as the quotient of $\mathcal{R}\mathbb{E}(X;A,B)$ by the
equivalence relation generated by unitary equivalence, translation of
$F$ along norm-continuous path, and the addition of degenerate
Kasparov cycles. There is an intersection product
$$
\mathcal{R}KK(X;A,B)\x \mathcal{R}KK(X;B,C)\raw \mathcal{R}KK(X;A,C),$$
enabling one to define a category \RKKX\  in the obvious way; the objects
are $C(X)$ \ca s, the arrows are the groups $\mathcal{R}KK(X;-,-)$,
composed through the intersection product. We are now in a position
to state 
\begin{Conjecture}
There is a functor from \Po\ to \RKKI\ that on objects defines a strict
quantization.
\end{Conjecture}


\begin{thebibliography}{99}
\itemsep=\smallskipamount

\bibitem{AS1} Atiyah, M.F.  and  Singer, I.M.: The index of elliptic operators I. 
\textit{Ann.\ Math.} \textbf{87} (1968), 485--530. 

\bibitem{BFFLS}   Bayen, F., Flato, M.,  Fronsdal, C.,  Lichnerowicz, A., and
Sternheimer, D.: Deformation theory and
quantization. I, II. \textit{
Ann. Phys.\ (N.Y.)} {\bf 110} (1978), 61--110, 111--151.

 \bibitem{Bla}  Blackadar, B.: \textit{K-theory for Operator Algebras}, 2nd ed. 
 Cambridge University Press, Cambridge, 1999.

 \bibitem{Bla2} Blanchard, E.: Tensor products of $C(X)$-algebras over $C(X)$.
\textit{Ast\'{e}risque}  \textbf{232} (1995),
81--92.

\bibitem{Bla1} Blanchard, E.: D\'{e}formations de $C^*$-alg\`{e}bres 
de Hopf. \textit{Bull.\ Soc.\ math.\ France} \textbf{124} (1996), 141--215.

\bibitem{BW} Bursztyn, H. and  Waldmann, S.: 
Algebraic Rieffel induction, formal Morita
equivalence, and applications to deformation quantization.
  \textit{J.\ Geom.\ Phys.} \textbf{37} (2001), 
307--364.

\bibitem{CW}  Cannas da Silva, A. and  Weinstein, A.: \textit{
 Geometric Models for Noncommutative Geometry}, Berkeley Mathematics
 Lecture Notes \textbf{10}, American Mathematical Society, Providence,
 RI, 1999.

 \bibitem{Con80} Connes, A.:
A survey of foliations and operator algebras. In:
Kadison, R. (ed.), 
\textit{Operator Algebras and Applications, Part I (Kingston, Ont., 1980), 
Proc.\ Sympos.\ Pure Math.}  \textbf{38}, Amer. Math. Soc., Providence, R.I., 1982,
pp. 521--628.

\bibitem{Con}  Connes, A.: \textit{Noncommutative Geometry},
Academic Press, San Diego, 1994.

\bibitem{CS} Connes, A. and  Skandalis, G.:
The longitudinal index theorem for foliations. \textit{Publ.\ Res.\ Inst.\
Math.\ Sci.} \textbf{20} (1984),  1139--1183. 

\bibitem{CDW} 
Coste, A.,  Dazord, P. and   Weinstein, A.:
Groupo\"{\i}des symplectiques.
 \textit{Publ.\ D\'{e}pt.\ Math.\ Univ. C.\
Bernard--Lyon I} \textbf{2A} (1987), 1--62.

\bibitem{Cou} 
Courant, T.J.:  Dirac manifolds. \textit{Trans.\ Amer.\ Math.\ Soc.}
\textbf{319} (1990),  631--661.

\bibitem{CM}
Crainic, M.  and Moerdijk, I.: A homology theory for \'{e}tale groupoids
\textit{J.\ Reine Angew.\ Math.} \textbf{521} (2000),   25--46. 

\bibitem{Dix} Dixmier, J.: \textit{$C^*$-Algebras},  North--Holland, Amsterdam, 1977.

\bibitem{Ech} 
Echterhoff, S., Kaliszewski, S., Quigg, J., and Raeburn, I.:
Naturality and induced representations. 
\textit{Bull.\ Austral.\ Math.\ Soc.}  \textbf{61} (2000), 415--438. 

\bibitem{ENN}   Elliott, G.A.,  Natsume, T., and  Nest, R.:
The Atiyah--Singer index theorem as passage to the classical limit in
quantum mechanics.  \textit{Commun.\ Math.\ Phys.} \textbf{182}  (1996),
505--533.

\bibitem{Fed}  Fedosov, B.: \textit{Deformation Quantization and Index Theory}.
Akademie--Verlag, Berlin, 1996.

 \bibitem{Fol}
Folland, G.B.:  \textit{Harmonic Analysis on Phase Space},  Princeton
University Press, Princeton, 1989.

\bibitem{Got1} Gotay,  M.J.: Functorial geometric quantization and Van Hove's theorem.
 \textit{
Internat.\ J.\ Theoret.\ Phys.} \textbf{19} (1980), 139--161.

\bibitem{Got2} Gotay, M. J.: On the Groenewold-Van Hove problem for $\mathbb{R}^{2n}$.
\textit{J.\  Math.\  Phys.} \textbf{40} (1999),  2107--2116.

\bibitem{GGT} Gotay, M. J.,  Grundling, H. B. and Tuynman, G. M.:
Obstruction results in quantization theory. 
\textit{J.\  Nonlinear Sci.} \textbf{6} (1996), 469--498. 

 \bibitem{Haa}  Haag, R.: \textit{Local Quantum Physics}, 2nd ed.,
 Springer, Heidelberg, 1996.

\bibitem{HR} Higson, N. and Roe, J.: \textit{Analytic K-homology}.
Oxford University Press, Oxford, 2000.

\bibitem{HS}
Hilsum, M. and    Skandalis, G.:  Morphismes $K$-orient\'{e}s d'espaces de feuilles et
fonctorialit\'{e} en th\'{e}orie de Kasparov (d'apr\`{e}s une conjecture d'A. Connes).
\textit{Ann.\ Sci.\ É\'{E}cole 
Norm.\ Sup.\ (4)}  \textbf{20} (1987),  325--390.

\bibitem{K1} 
Karasev, M.V.: The Maslov quantization conditions in higher cohomology and analogs
 of notions developed in Lie theory for canonical fibre bundles of
 symplectic manifolds.  I, II.
  \textit{Selecta Math. Soviet.} \textbf{8} (1989),
  213--234, 235--258.

\bibitem{Kas0} Kasparov, G. G.: The operator $K$-functor and extensions of $C^*$-algebras.
Math.\ USSR Izvestija  SSSR \textbf{16} (1981), 513--572.

\bibitem{Kas1} Kasparov, G. G.: Equivariant $KK$-theory and the Novikov conjecture.
\textit{Invent.\ Math.}  \textbf{91} (1988), 147--201.

\bibitem{KW} Kirchberg, E. and Wassermann, S.: Operations on continuous bundles of 
$C^*$-algebras. \textit{Math.\ Ann.} \textbf{303} (1995), 677--697.

\bibitem{KL3} Klimek, S. and   Lesniewski, A.:   Quantum Riemann surfaces for arbitrary Planck's
constant. \textit{J.\ Math.\ Phys.} \textbf{37}
 (1996),  2157--2165.

\bibitem{Lance} Lance, E.C.:  \textit{Hilbert
 $C^*$-Modules. A Toolkit for Operator Algebraists. LMS Lecture Notes}
  \textbf{210}.  Cambridge University Press, Cambridge, 1995.

\bibitem{NPL1} Landsman,  N.P.:  Rieffel induction as 
generalized quantum Marsden-Weinstein reduction.
\textit{J.\ Geom.\ Phys.}  \textbf{15} (1995), 285--319, Err.\ ibid.\ \textbf{17} (1995), 298;
eprint \texttt{hep-th/9305088}.  

\bibitem{NPL3}  Landsman, N.P.:
 \textit{Mathematical Topics Between Classical
and Quantum Mechanics}.
Springer, New York, 1998.

\bibitem{LGCA}  Landsman, N.P.: Lie groupoid $C^*$-algebras and Weyl quantization.
 \textit{Commun.\  Math.\ Phys.} \textbf{206} (1999), 367--381;
 eprint \texttt{math-ph/9807028}.

\bibitem{NPLMRW} Landsman, N.P.: The Muhly-Renault-Williams 
theorem for Lie groupoids and its classical counterpart.
\textit{Lett.\ Math.\ Phys.} \textbf{54} (2000), 43--59;
 eprint \texttt{math-ph/0008005}.

\bibitem{OBWF} Landsman, N.P.: Quantized reduction as a tensor product. In: 
 Landsman, N.P.,  Pflaum, M.  and  Schlichenmaier, M.  (eds),
\textit{Quantization of Singular Symplectic Quotients},
Birkh\"{a}user, Basel, 2001, pp.\ 137--180; eprint \texttt{math-ph/0008004}. 

\bibitem{FM} Landsman, N.P.:   Operator algebras and
Poisson manifolds associated to  groupoids.
\textit{Commun.\ Math.\ Phys.} \textbf{222} (2001), 97--116; eprint \texttt{math-ph/0008036}.

\bibitem{LR}  Landsman,  N.P. and  Ramazan, B.:  Quantization of Poisson 
algebras associated to  Lie algebroids. 
\textit{Contemp.\ Math.} \textbf{282}(2001), 159--192; eprint \texttt{math-ph/0001005}.
 
\bibitem{Lang} Lang, S.: \textit{Algebra, 3d ed.}, Addison--Wesly, Reading, 1993.

 \bibitem{LG} Le Gall, P.-Y.: Th\'{e}orie de Kasparov \'{e}quivariante et groupo\"{\i}des. I.
 \textit{K-Theory} \textbf{16} (1999), 361--390.

\bibitem{Mac} 
Mackenzie, K.C.H.: \textit{Lie Groupoids and Lie Algebroids in
Differential Geometry},  Cambridge University Press, Cambridge, 1987.

\bibitem{MX} Mackenzie, K.C.H., Xu, P.: Integration of Lie bialgebroids. \textit{Topology}
\textbf{39} (2000), 445--476.

\bibitem{Moe}  Moerdijk,  I.: Toposes and groupoids. 
 \textit{Lecture Notes in Math.}  \textbf{1348} (1988)  280--298.

\bibitem{MM} 
 Moerdijk, I. and  Mr\v{c}un, J.:  On integrability of infinitesimal
actions. eprint \texttt{math.DG/0006042}.

\bibitem{Mrc1}   Mr\v{c}un, J.: \textit{Stability and Invariants of
Hilsum--Skandalis Maps},  Ph.D. thesis, University of Utrecht, 1996.

\bibitem{Mrc2} Mr\v{c}un, J.:  Functoriality of the bimodule associated to a 
Hilsum--Skandalis map. \textit{$K$-Theory}  \textbf{18} (1999),  235--253.

\bibitem{MRW}  Muhly, P., Renault, J.  and  Williams, D.:
Equivalence and 
isomorphism for groupoid \ca s.
\textit{J.\ Operator Th.} \textbf{17} (1987), 3--22.

\bibitem{Nil} Nilsen, M.: $C^*$-bundles and $C_0(X)$-algebras.
\textit{Indiana Univ.\ Math.\ J.} \textbf{45} (1996), 463--477.

\bibitem{PT} Park, E. and Trout, J.:  Representable $E$-Theory for
$C_0(X)$-Algebras. \textit{J.\ Funct.\ Anal.} \textbf{177} (2000), 178--202.

\bibitem{Ram}  Ramazan, B.:  \textit{
Deformation Quantization of Lie--Poisson Manifolds}, Ph.D.\ thesis,
Universit\'{e} d'Orl\'{e}ans, 1998.

\bibitem{Ren}
 Renault, J.: \textit{A Groupoid Approach to $C^*$-algebras}, \textit{
Lecture Notes in Mathematics} \textbf{793}, Springer, Berlin, 1980.

\bibitem{Rie1} Rieffel,   M.A.: Rieffel, M.A.: Induced representations of
$C^*$-algebras.
 \textit{Adv.\ Math.} \textbf{13} (1974), 176--257.

\bibitem{Rie2}    Morita equivalence for $C^*$-algebras and
$W^*$-algebras.
 \textit{J.\ Pure Appl.\ Alg.} \textbf{5} (1974), 51--96.

\bibitem{Rie3} Rieffel, M.A.: Deformation quantization of Heisenberg manifolds. 
\textit{Commun.\ Math.\ Phys.}  \textbf{122} (1989), 531-562.

\bibitem{Rie4}  Rieffel, M.A.: Noncommutative tori--a case study of noncommutative
differentiable manifolds. \textit{Contemp.\ Math.} \textbf{105} (1990), 191--211.
 

\bibitem{RS} Rieffel, M.A. and Schwarz, A.: 
Morita equivalence of multidimensional noncommutative tori.
 \textit{Internat.\ J.\ Math.} \textbf{10} (1999), 289--299. 
 

\bibitem{Sch} Schweizer, J.: \textit{Crossed products by equivalence
bimodules}, Univ.\ T\"{u}bingen preprint (1999).

\bibitem{Vor} Voronov, T.: Quantization on supermanifolds and the analytic proof of the
Atiyah--Singer index theorem.  \textit{J.\ Soviet Math.} \textbf{64}  (1993), 993-1069.

\bibitem{W1} Weinstein, A.: The
 local structure of Poisson manifolds. \textit{J.\ Diff.\ Geom.}  \textbf{
 18} (1983), 523--557,  Err.\ \textit{ibid.}\ \textbf{22} (1985), 255.

 \bibitem{Xu2}  Xu, P.:  Morita equivalence of Poisson manifolds.
 \textit{Commun.\ Math.\ Phys.} \textbf{142} (1991), 493--509.
\end{thebibliography}
 \end{document}